\documentclass[10pt, paper=a4, UKenglish
]{article}
\usepackage{graphicx}
\def\Title#1{\begin{center} {\Large #1 } \end{center}}
\def\Author#1{\begin{center}{ \sc #1} \end{center}}
\def\Address#1{\begin{center}{ \it #1} \end{center}}

\newcommand\pubblock{\rightline{\begin{tabular}{l} Proceedings of the CTD/WIT 2019\\ \pubnumber\\
         \pubdate  \end{tabular}}}

\newenvironment{Abstract}{\begin{quotation} \begin{center} 
             \large ABSTRACT \end{center}\bigskip 
      \begin{center}\begin{large}}{\end{large}\end{center} \end{quotation}}

\newenvironment{Presented}{\begin{quotation} \begin{center} 
             PRESENTED AT\end{center}\bigskip 
      \begin{center}\begin{large}}{\end{large}\end{center} \end{quotation}}

\def\beq{\begin{equation}}
\def\eeq#1{\label{#1}\end{equation}}
\def\eeqn{\end{equation}}

\def\beqa{\begin{eqnarray}}
\def\eeqa#1{\label{#1}\end{eqnarray}}
\def\eeqan{\end{eqnarray}}

\let\bar=\overbar

\def\Dslash{\not{\hbox{\kern-4pt $D$}}}
\def\dslash{\not{\hbox{\kern-2pt $\del$}}}

\def\msb{{\bar{\ssstyle M \kern -1pt S}}}

\usepackage{color}
\usepackage{lineno}
\usepackage{subfig}
\usepackage{hyperref}
\usepackage[dvipsnames]{xcolor}
\usepackage{dcolumn}
\usepackage{tikz}
\usepackage{amsmath,amsfonts,amsthm,bm}

\newcommand\pubnumber{PROC-CTD19-098}

\newcommand\pubdate{\today}

\def\affiliation{
On behalf of Belle II collaboration, \\
$^\mathrm{*}$Charles University, Czech Republic \\
$^\mathrm{\dagger}$High Energy Accelerator Research Organization, Japan \\
$^\mathrm{\ddagger}$Tokyo Institute of Technology, Japan \\ 
$^\mathrm{\star}$The Graduate University for Advanced Studies, Japan \\
$^\mathrm{\amalg}$German Electron Synchrotron, Germany
}

\newcommand{\conference}{Connecting the Dots and Workshop on Intelligent Trackers (CTD/WIT 2019)\\
Instituto de F\'isica Corpuscular (IFIC), Valencia, Spain\\ 
April 2-5, 2019}

\usepackage{fancyhdr}
\definecolor{mygrey}{RGB}{105,105,105}
\begin{document}


\large
\begin{titlepage}
\pubblock

\vfill
\Title{Calibration and alignment of the Belle II tracker}
\vfill

\Author{Jakub Kandra$^\mathrm{*}$, Tadeas Bilka$^\mathrm{*}$, Lucia Kapitanova$^\mathrm{*}$, Makoto Uchida$^\mathrm{\ddagger}$, \\ Hitoshi Ozaki$^\mathrm{\dagger}$, Thanh Van Dong$^\mathrm{\star}$ and Claus Kleinwort$^\mathrm{\amalg}$}
\Address{\affiliation}
\vfill

\begin{Abstract}
The physics goals the Belle II experiment require an exceptionally good alignment of all the components of the Belle II tracker. The Belle II tracker is composed of the DEPFET based pixel silicon detector, four layers of double sided silicon strip detector, a low material budget drift chamber, all three operating in a solenoidal 1.5 T B field, which is affected by the final focusing system of the accelerator. Each component of these three components must be aligned with an accuracy significantly better than the point resolution of the detector that for the PXD is order of 10 microns. The Belle II alignment software is based on the Millepede II package and uses cosmics and collision data to constrain the weak modes. The performance of the alignment algorithms was tested on the phase 2 collision data collected during spring 2018. Good alignment of the vertex detector was essential to demonstrate the nano-beam collision scheme of the accelerator and check the quality of the impact parameter resolution, which is essential for time-dependent CP violation studies at the B factory.
\end{Abstract}

\vfill

\begin{Presented}
\conference
\end{Presented}
\vfill
\end{titlepage}
\def\thefootnote{\fnsymbol{footnote}}
\setcounter{footnote}{0}
%

\normalsize 


\section{Introduction}
The Belle II detector is an upgrade of one of the famous B factories from beginning of $\mathrm{21^{th}}$ century. The detector is built on asymmetric electron-positron accelerator, SuperKEKB, in Japanese High Energy Accelerator Research Organization in Tsukuba. The new accelerator is designed to produce $\mathrm{\sim 50 \ ab^{-1}}$ collisions and provide clean environment for production of B meson pairs via $\mathrm{\Upsilon(4S)}$ resonance decay. \\
Expanding four layers of strip sensors by two additional inner layers of pixel sensors and adding more layers of wires in central drift chamber are the main upgrades of Belle II tracker~\cite{DesignReport}. The first possibility to integrate upgraded subsystems and test cooperation with data acquisition system of the Belle II detector was during Belle II Commissioning run. \\
For first collisions in April 2018, determination of alignment and calibration corrections of sensors and wires was necessary. For this purpose Millepede II algorithm is implemented into alignment and calibration procedure~\cite{MillepedeII}. We describe here the procedure to provide precise alignment and new technique to validate and monitor alignment parameters. We present precision and stability of the alignment too.

\section{Alignment parametrization of the vertex detector}
\begin{figure}[!htb]
  \begin{minipage}[c]{0.45\linewidth}
  \includegraphics[width=1.0\linewidth]{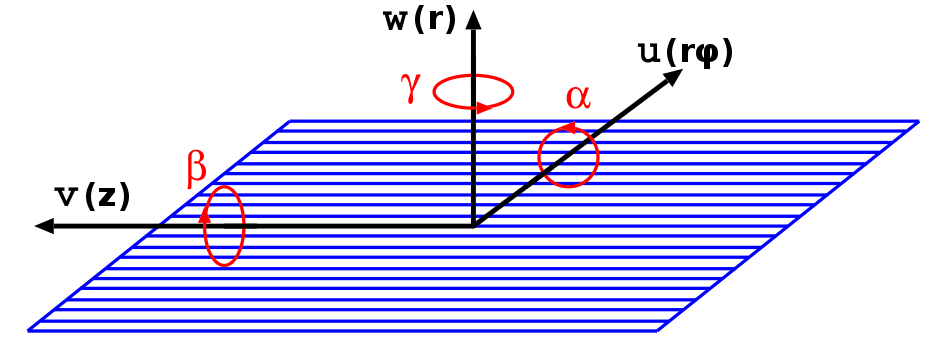}
  \end{minipage}
  \begin{minipage}[c]{0.55\linewidth}
  \caption{Rigid body parameters in local coordinate system of sensor: Axis $\mathrm{u}$ in direction of shorter side of sensor with rotation $\mathrm{\alpha}$ around itself, axis $\mathrm{v}$ in direction of longer side of sensor with rotation $\mathrm{\beta}$ around itself and axis $\mathrm{w}$ in direction of perpendicular to plane of sensor with rotation $\mathrm{\gamma}$ around itself.}
  \label{fig:RigidBodyParameters}
  \end{minipage}  
\end{figure}

The rigid body parameters (Figure~\ref{fig:RigidBodyParameters}) are three parameters for description of relative movements of sensors and three parameters for relative rotations. They are defined as corrections applied during transformations of measured hit positions from local coordinate system of sensor to global coordinate system. The global coordinate system is used for parametrization of reconstructed tracks. The standard labels of alignment parameters are $u, \ v \ and \ w$ for shifts of sensors and $\alpha, \ \beta \ and \ \gamma$ for rotations of sensors around axis of local coordinates. 

\begin{figure}[!htb]
  \centering
  \begin{minipage}[c]{0.42\linewidth}
  \includegraphics[width=1.0\linewidth, trim={0cm, 0cm, 0cm, 1.0cm}, clip]{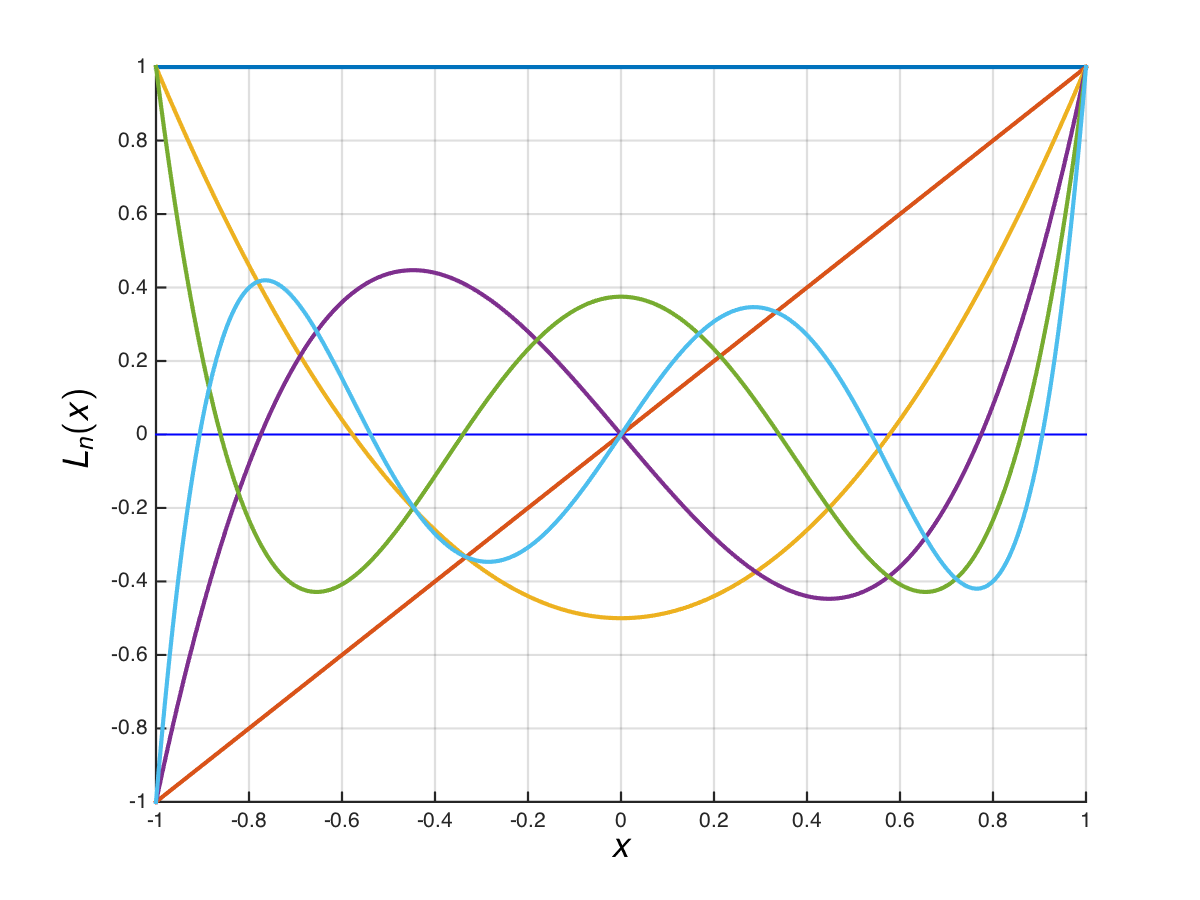} 
  \vspace{-5.4cm} \\
  \text{\fontsize{8}{0} \selectfont \hspace{1.0cm} \textcolor{NavyBlue}{ $\boldsymbol{L_0(x) = 1}$}, \hspace{1.2cm} \textcolor{RedOrange}{$\boldsymbol{L_1(x) = x}$},} \\
\text{\fontsize{8}{0} \selectfont \hspace{1.1cm} \textcolor{Dandelion}{$\boldsymbol{L_2(x) = \frac{1}{2} (3 x^2 - 1)}$},} \\
\text{\fontsize{8}{0} \selectfont \hspace{1.22cm} \textcolor{Plum}{$\boldsymbol{L_3(x) = \frac{1}{2} (5 x^3 - 3x)}$},} \vspace{2.1cm} \\
\text{\fontsize{8}{0} \selectfont \hspace{1.8cm} \textcolor{ForestGreen}{$\boldsymbol{L_4(x) = \frac{1}{8} (35x^4 - 30x^2 + 3)}$},} \\
\text{\fontsize{8}{0} \selectfont \hspace{1.5cm}\textcolor{ProcessBlue}{$\boldsymbol{L_5(x) = \frac{1}{8} (63x^5 - 70x^3 + 15x)}$}} \\  
  \end{minipage}
  \begin{minipage}[c]{0.55\linewidth}
  \includegraphics[width=0.32\linewidth]{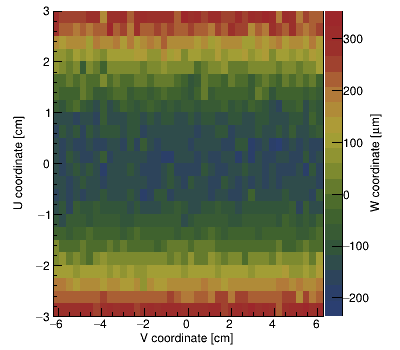}
  \includegraphics[width=0.32\linewidth]{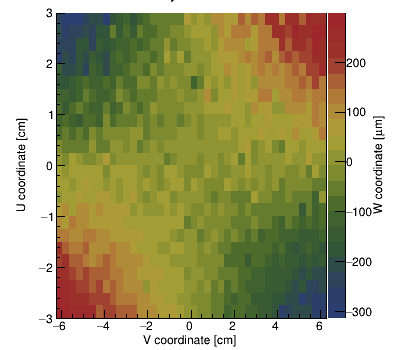}
  \includegraphics[width=0.32\linewidth]{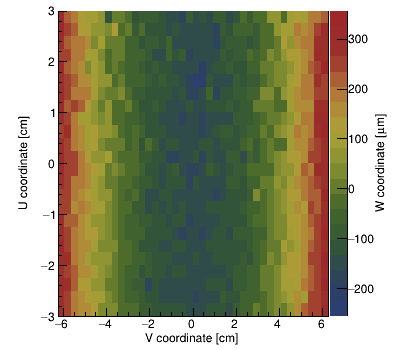} 
  \vspace{-1.9cm} \\
  \text{\fontsize{18}{18} \selectfont \hspace{0.65cm} \textcolor{White}{$\boldsymbol{P_{20}}$} \hspace{1.75cm} \textcolor{White}{$\boldsymbol{P_{11}}$} \hspace{1.75cm} \textcolor{White}{$\boldsymbol{P_{02}}$}} 
  \vspace{0.1cm} \\
  \text{\fontsize{9}{9} \selectfont \hspace{0.25cm} \textcolor{White}{$L_2(u) \cdot L_0(v)$} \hspace{1.0cm} \textcolor{White}{$L_1(u) \cdot L_1(v)$} \hspace{1.0cm} \textcolor{White}{$L_0(u) \cdot L_2(v)$} } 
  \vspace{0.89cm} \\  
  \includegraphics[width=0.24\linewidth]{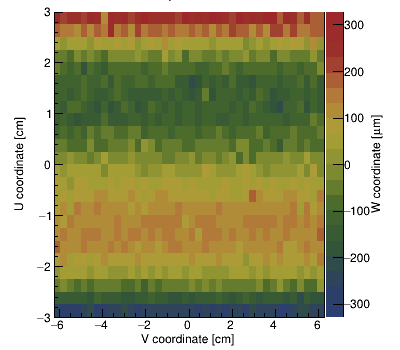}
  \includegraphics[width=0.24\linewidth]{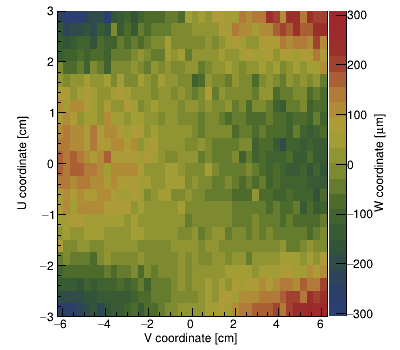}
  \includegraphics[width=0.24\linewidth]{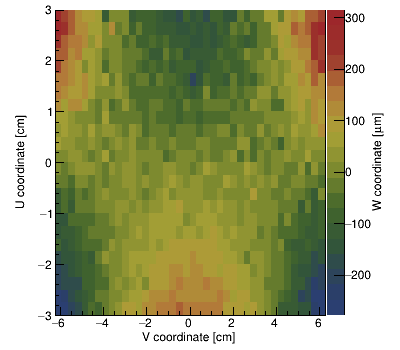}
  \includegraphics[width=0.24\linewidth]{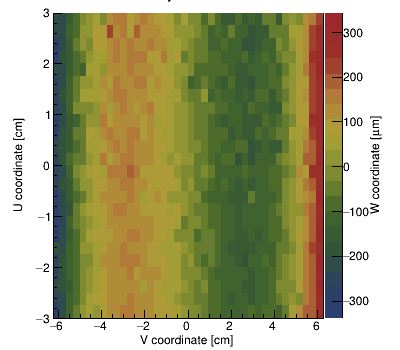} 
  \vspace{-1.88cm} \\
  \text{\fontsize{18}{0} \selectfont \hspace{0.3cm} \textcolor{White}{$\boldsymbol{P_{30}}$} \hspace{1.0cm} \textcolor{White}{$\boldsymbol{P_{21}}$} \hspace{1.0cm} \textcolor{White}{$\boldsymbol{P_{12}}$} \hspace{1.0cm} \textcolor{White}{$\boldsymbol{P_{03}}$}} 
  \vspace{0.0cm} \\
  \text{\fontsize{5}{0} \selectfont \hspace{0.1cm} \textcolor{White}{$L_3(u) \cdot L_0(v)$} \hspace{0.68cm} \textcolor{White}{$L_2(u) \cdot L_1(v)$} \hspace{0.66cm} \textcolor{White}{$L_1(u) \cdot L_2(v)$} \hspace{0.66cm} \textcolor{White}{$L_0(u) \cdot L_3(v)$}} 
  \vspace{0.2cm} \\  
  \end{minipage}  
\caption{Legendre polynomials (left) and surface deformation parametrization (right) using polynomials}
  \label{fig:LegendreParameters} 
\end{figure}

\noindent
The surface deformation parameters (Figure~\ref{fig:LegendreParameters}) are used for description of surface properties of sensor. With elimination surface deformations of the sensors we are able to improve hit position uncertainty. In local coordinate system, the surface is parametrized as $\mathrm{w \ = z w(u,v)}$ using Legendre polynomials. Advantage of using Legendre polynomials is their orthogonality $x \in [-1, +1]: \ \int^{+1}_{-1} L_i \cdot L_j \sim \delta_{ij} (= 0 \ \mathrm{for} \ i \neq j)$.  If a sensor illumination is uniform at least along one coordinate, the contributions from different orders are independent.
\section{Vertex detector alignment validation}
\begin{figure}[!htb]
  \includegraphics[width=1.0\linewidth]{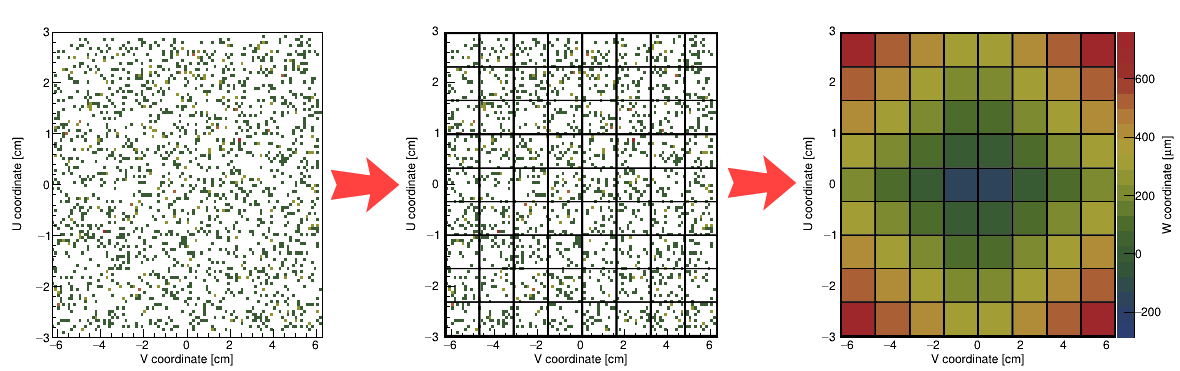}
  \caption{Estimation of $w$ coordinate proceed in three steps. We divide sensor's surface to m $\times$ n matrix (center). Average value of the w residual is then estimated for each cell of the matrix with the use of formulas~\ref{eq:estimationWCoordinate} for all hits in cell. Each contribution to averaged value is weighted by squared slopes of track in sensor (right). The estimation should be done for both measurement directions.}
  \label{fig:PlanarityExplanation}
\end{figure}

Monitoring of alignment parameters can be done in different ways. One of useful methods is too monitor quality of reconstructed data per each sensor. Standard validation method is monitoring of track-to-hit residual distributions. For surface deformations we monitor monitor residual distributions in $w$ coordinate. However measurement is provided in $u$ and $v$ coordinate. For small deformations the $w$ coordinate is estimated from measurements as:
\begin{equation}
  r_{W} = \frac{r_{U} }{\tan{\alpha_U}} \quad or \quad \small r_{W} = \frac{r_{V}}{\tan{\alpha_V}},
  \label{eq:estimationWCoordinate}
\end{equation}
where $r_{U, V, W}$ are residuals and $\tan{\alpha_{U,V}}$ are slopes of track in sensor. This is illustrate in Figure~\ref{fig:PlanarityExplanation}.

\begin{figure}
 \centering
 \includegraphics[width=0.31\linewidth, trim={0mm, 0mm, 0mm, 7mm}, clip]{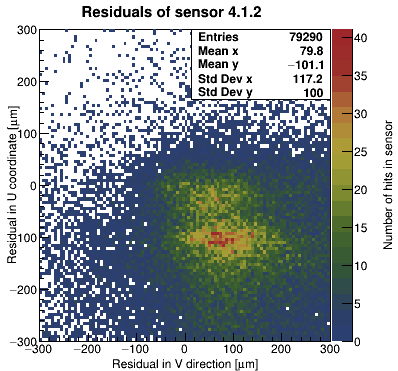}
 \includegraphics[width=0.31\linewidth, trim={0mm, 0mm, 0mm, 7mm}, clip]{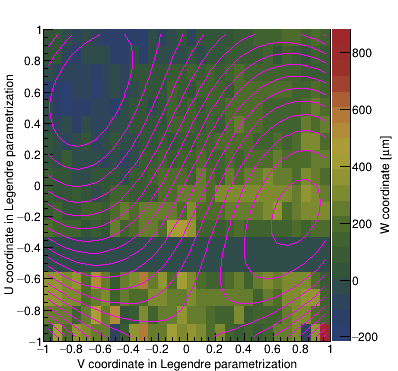}
 \includegraphics[width=0.31\linewidth, trim={0mm, 0mm, 0mm, 7mm}, clip]{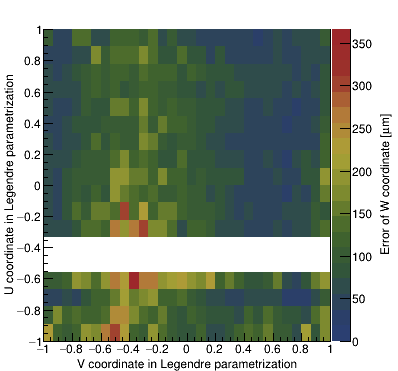}

 \includegraphics[width=0.31\linewidth, trim={0mm, 0mm, 0mm, 7mm}, clip]{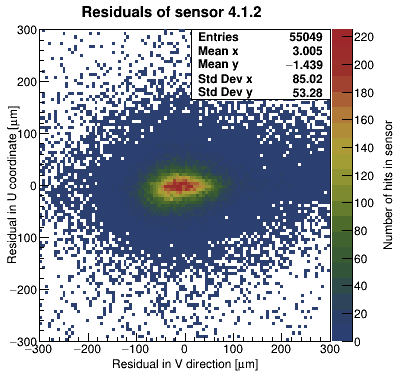}
 \includegraphics[width=0.31\linewidth, trim={0mm, 0mm, 0mm, 7mm}, clip]{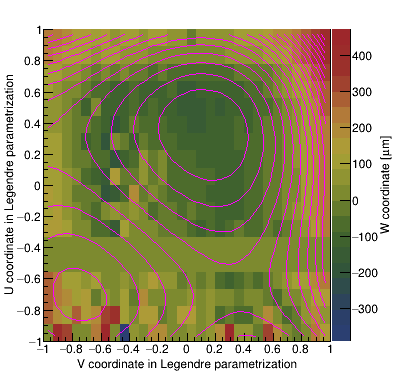}
 \includegraphics[width=0.31\linewidth, trim={0mm, 0mm, 0mm, 7mm}, clip]{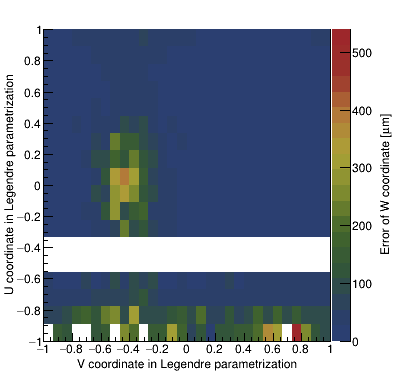}
 
 \includegraphics[width=0.31\linewidth, trim={0mm, 0mm, 0mm, 7mm}, clip]{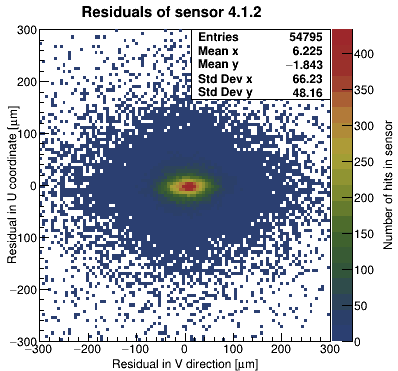}
 \includegraphics[width=0.31\linewidth, trim={0mm, 0mm, 0mm, 7mm}, clip]{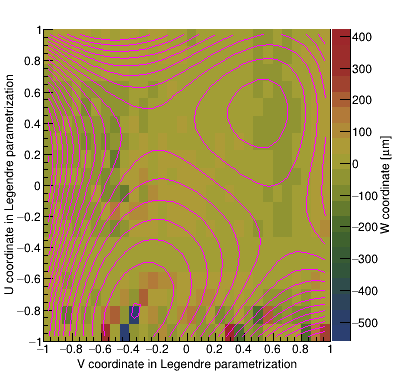}
 \includegraphics[width=0.31\linewidth, trim={0mm, 0mm, 0mm, 7mm}, clip]{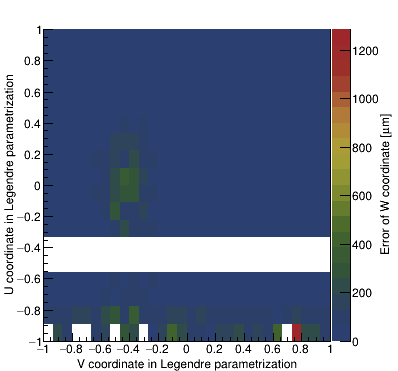}
 \caption{Alignment validation plots for sensor 4.1.2: Histograms filled before application alignment constants (top row), after application of rigid body (center row) and surface parameters $\mathrm{P_{20}, P_{11} \ and \ P_{02}}$ (bottom row), we refer to these parameters as ''Simple surface''. Plots are residual distributions for $u$ and $v$ coordinates (left column), distribution for $w$ coordinate (center column) and $w$ coordinate error distribution (right column).}
 \label{fig:Sensor4-1-2} 
\end{figure}

\noindent
The alignment parameters can be estimated from validation procedure from the residuals plots~\ref{fig:Sensor4-1-2}. The shifts in $u$ and $v$ direction can be estimated as means of distributions. Other alignment parameters can be determined by fitting surface validation plot~\ref{fig:Sensor4-1-2}. The fitting procedure is done in two steps: transformation from a sensor’s local coordinate system to Legendre system and fitting by 2D Legendre polynomial function:
\begin{align}
  w (u,v) & = P_{W} \cdot L_0(u) \cdot L_0(v) + P_{\alpha} \cdot L_0(u) \cdot L_1(v) + P_{\beta} \cdot L_1(u) \cdot L_0 (v) \ + \nonumber \\
  & + P_{20} \cdot L_2(u) \cdot L_0(v) + P_{11} \cdot L_1(u) \cdot L_1(v) + P_{02} \cdot L_0(u) \cdot L_2 (v) \ + \\
  & + P_{30} \cdot L_3(u) \cdot L_0(v) + P_{21} \cdot L_2(u) \cdot L_1(v) + P_{12} \cdot L_1(u) \cdot L_2 (v) \ + P_{03} \cdot L_0(u) \cdot L_3 (v) \nonumber  
\label{eq:validationEquation}
\end{align}
\noindent
where $\mathrm{L_N}$ are Legendre polynomials same as in Figure~\ref{fig:LegendreParameters}, $\mathrm{P_M}$ are alignment corrections. One of the biggest disadvantages of this method is missing possibility to estimate $\mathrm{P_{\gamma}}$ parameter associated with angle $\mathrm{\gamma}$.

\begin{table}
 \begin{minipage}[c]{0.6\linewidth}
 \centering
 \begin{tabular}{c D{|}{ \ \pm \ }{4.4} D{|}{ \ \pm \ }{4.4} D{|}{ \ \pm \ }{4.4}}
 P [$\mu$m] & \multicolumn{1}{c}{\emph{Before alignment}} & \multicolumn{1}{c}{\emph{Rigid  body}} & \multicolumn{1}{c}{\emph{Simple surface}} \tabularnewline
 \hline
 $P_{U}$      &  -94.59|0.56 &  -0.45|0.24 &  -1.77|0.22 \tabularnewline
 $P_{V}$      &   78.31|0.55 &   2.92|0.37 &   6.18|0.29 \tabularnewline
 $P_{W}$      &  119.93|0.52 &  12.46|0.25 &   1.43|0.16 \tabularnewline
 $P_{\alpha}$ &   87.56|0.98 &   4.74|0.60 &  -0.49|0.24 \tabularnewline
 $P_{\beta}$  &  -94.81|0.86 &  -1.88|0.53 &  -0.07|0.27 \tabularnewline
 $P_{02}$     &    4.97|1.12 &  96.00|0.78 &  -8.59|0.32 \tabularnewline
 $P_{11}$     &   69.98|1.53 &  37.26|0.90 &   5.04|0.46 \tabularnewline
 $P_{20}$     &  -12.64|1.03 &  60.28|0.60 &   4.12|0.35 \tabularnewline
 $P_{03}$     &  -46.88|1.00 &  51.79|0.93 &   8.72|0.39 \tabularnewline
 $P_{12}$     &    9.53|1.74 &  59.64|1.37 &   7.09|0.55 \tabularnewline
 $P_{21}$     & -118.53|1.65 & -10.12|1.30 & -12.86|0.63 \tabularnewline
 $P_{30}$     &   12.12|1.07 &  66.93|0.81 &   7.49|0.41 \tabularnewline
 \hline
 \hline
 \emph{$\mathrm{\sigma_{plot}}$} & \multicolumn{1}{c}{77.82} & \multicolumn{1}{c}{19.54} & \multicolumn{1}{c}{17.88}
 \end{tabular}
 \end{minipage}
  \begin{minipage}[c]{0.4\linewidth}
 \caption{Determination of alignment parameters using validation plots: Left column presents labels of determined parameters and other columns present results for each of the presented scenario in Figure~\ref{fig:Sensor4-1-2}. The last line represents uncertainties for each of the scenario. The uncertainty is determined as an error of $\sigma_{plot}$ from fitting of $\sigma_{plot} \cdot L_0(u) \cdot L_0(v)$ to the surface validation plot after application of the alignment corrections.}
\label{tab:Sensor-4-1-2}
\end{minipage}  
\end{table}

\section{Precision and stability of vertex detector alignment}
\begin{figure}[!htb]
  \centering
  \includegraphics[width=0.31\linewidth, trim={0mm, 0mm, 365mm, 31mm}, clip]{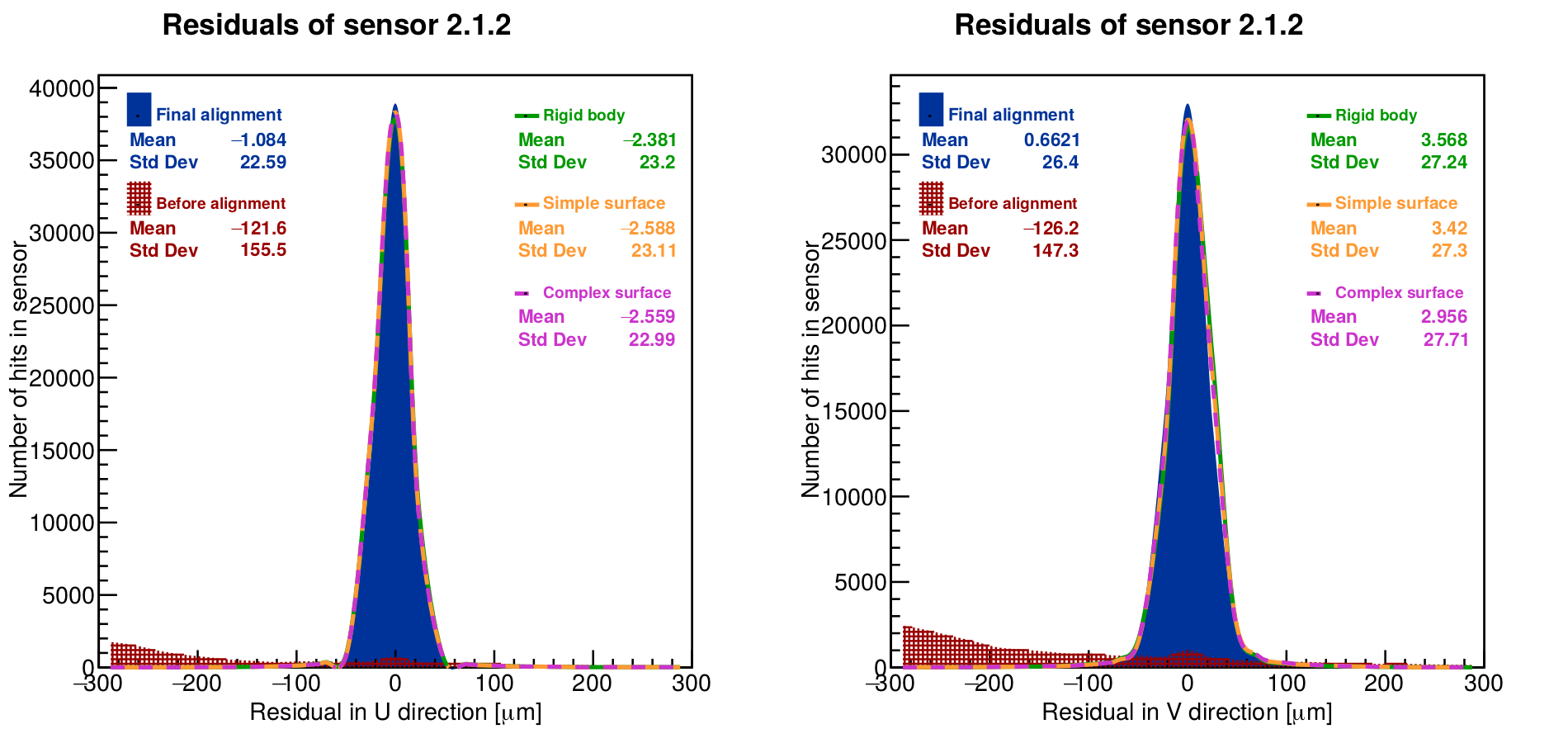}
  \includegraphics[width=0.31\linewidth, trim={0mm, 0mm, 365mm, 31mm}, clip]{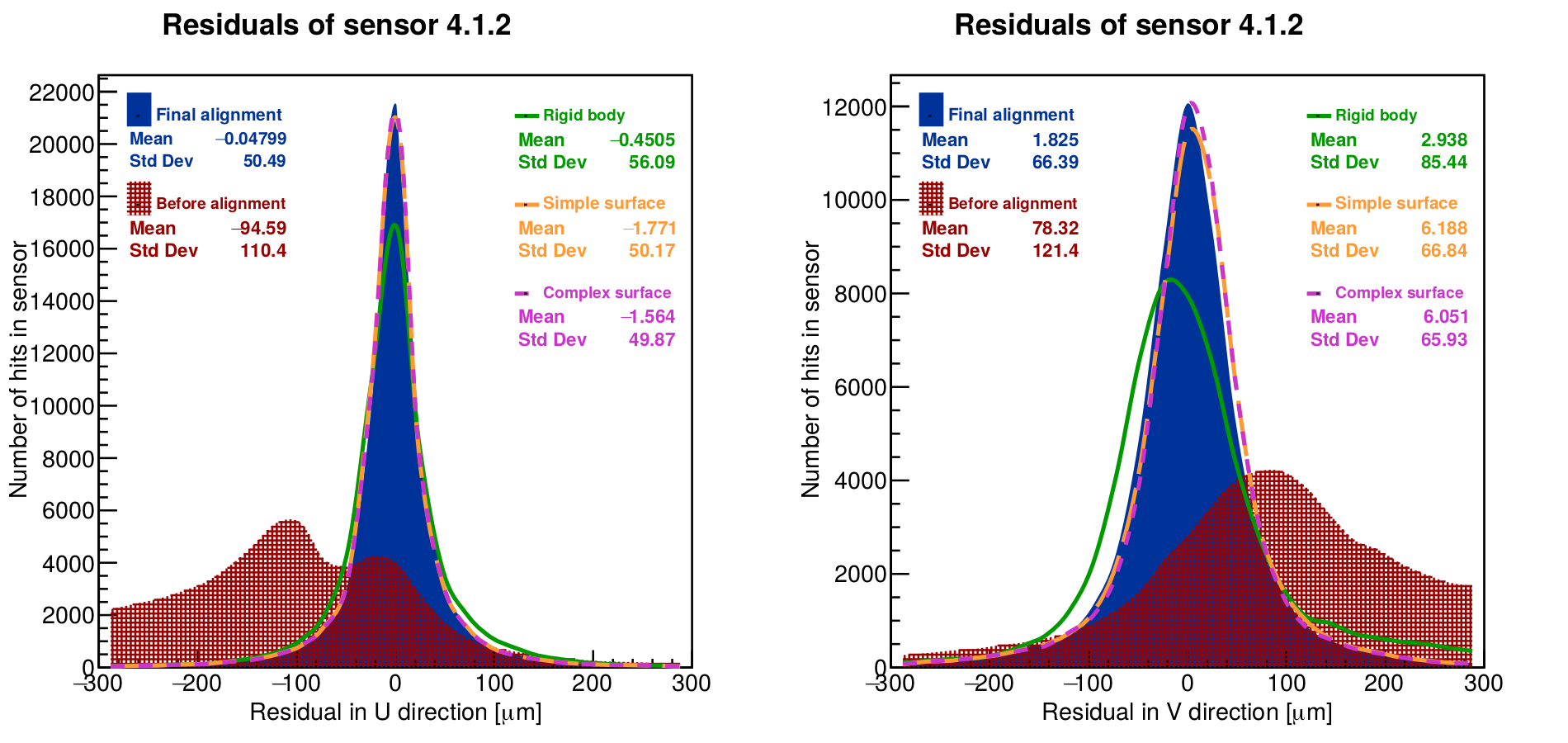}
  \includegraphics[width=0.31\linewidth, trim={340mm, 0mm, 25mm, 31mm}, clip]{residual-4-1-2.png}
\caption{Track-to-hit residual distributions in $u$ direction for pixel sensor 2.1.2 (left), strip sensor 4.1.2 (center) and in $v$ direction for same strip sensor (right).}
  \label{fig:PrecisionResiduals} 
\end{figure}

\begin{figure}[!htb]
\centering
\includegraphics[width=0.92\linewidth, trim={0mm, 0mm, 0mm, 0mm}, clip]{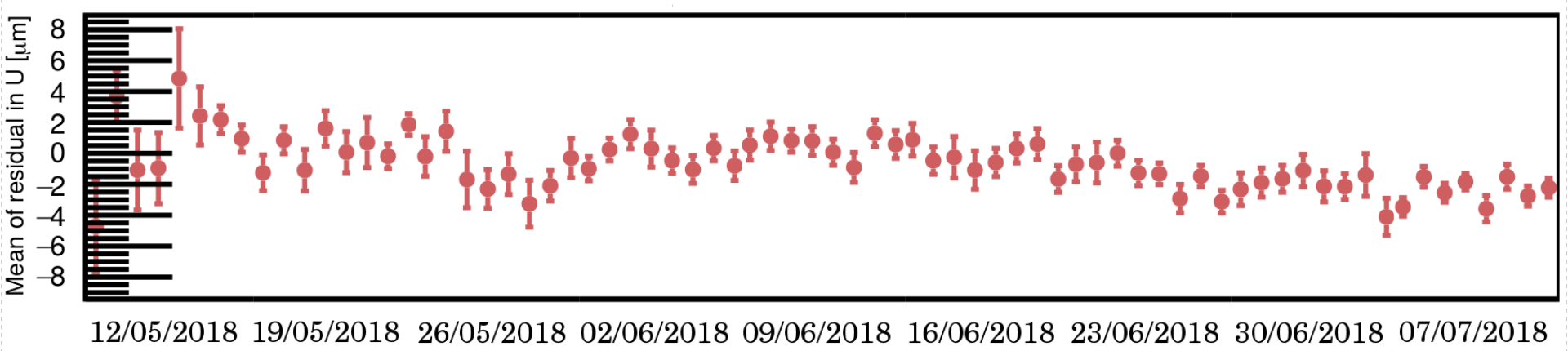}
 \caption{Stability measurement of sensor 4.1.2 during Commissioning run in 2018: Fluctuation of means of residuals in U direction as function of time is shown.}
 \label{fig:StabilityResiduals} 
\end{figure}

Precision of hit position measurement can be estimated from track-to-hit residual distributions. These distributions for selected pixel and strip sensor as function of applying alignment scenario are shown in Figure~\ref{fig:PrecisionResiduals}. ''Final alignment'' scenario is repetition of ''Simple surface'' scenario in two iterations to account for small corrections and non-linearities. The stability plot (Figure~\ref{fig:StabilityResiduals}) shows evolution of alignment parameter as function of time. 

\vspace{-2mm} 
\section{Calibration and alignment of central drift chamber}
The central drift chamber (CDC) is a main tracking detector in Belle II which consists of 56 sensitive layers (9 super layers), 
measuring a transverse momentum of charged particles, $dE/dx$ to provide the particle identification combined with measurement of other sub detectors.\\

\noindent
Since the displacement effects of wire positions at the end plates of each side mainly due to the deformation of end plates caused by the tension of wires after stringing were measured 
by the mechanical survey before the installation of CDC into Belle II detector, these effects are properly reflected in the reconstruction.
Further alignment of CDC was performed wire-by-wire using cosmic data samples with and without magnetic field.
Detailed procedures are described in Ref. \cite{thanh2019}. \\

\begin{figure}[!htb]
\begin{minipage}[c]{0.51\linewidth}
\includegraphics[width=\linewidth,  trim={8mm, 8mm, 5mm, 12mm}, clip]{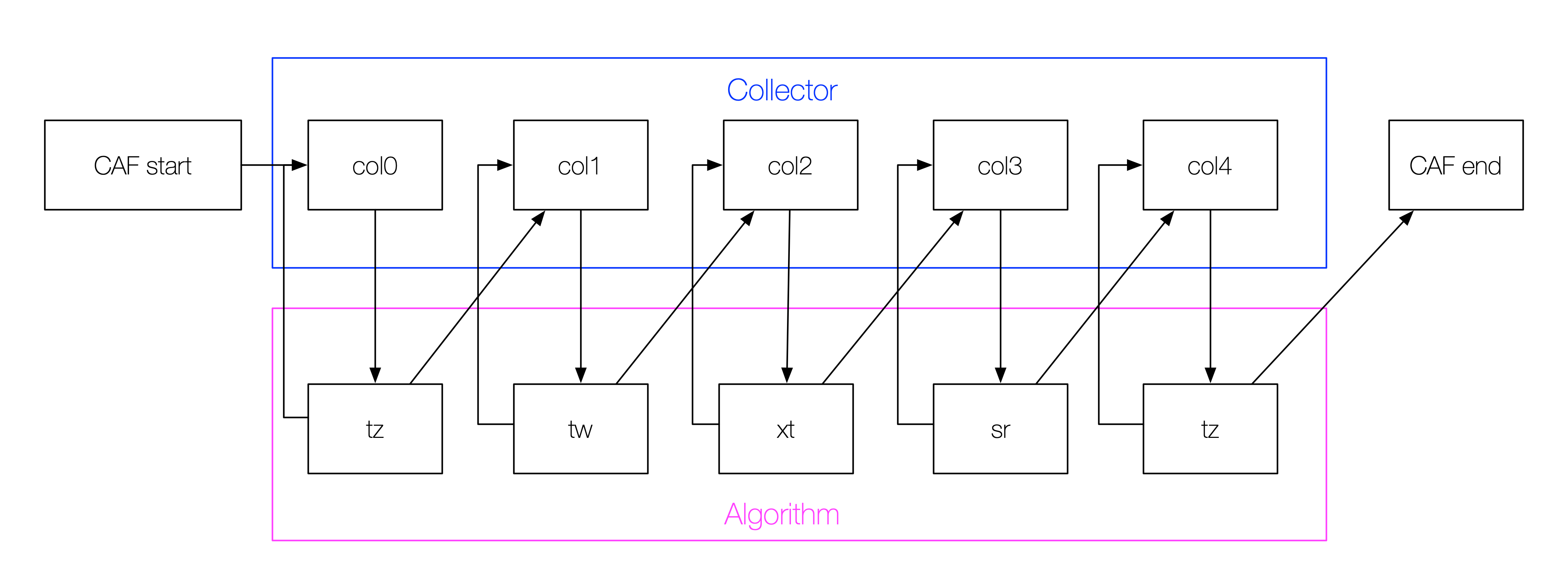}
\end{minipage}
\begin{minipage}[c]{0.49\linewidth}
\caption{Flowchart of CDC calibration. col$i(i=0,1,2,3,4)$, tz, tw, xt and sr mean collector process, time zero, time walk, XT, spatial resolution calibration algorithm, respectively.
All procedures described in the schematic view such as iteration and transition of state are controlled by a single CAF process. }
\label{fig:wf.cdc.calib}
\end{minipage}
\end{figure}

\noindent
The calibration of CDC proceeds to determine time zero ($T_0$), time walk, space-time relation (XT), and spatial resolution sequentially and iteratively. The schematic work flow is presented in Figure \ref{fig:wf.cdc.calib}. All procedures are implemented in the calibration and alignment framework (CAF) in Belle II software. 

\noindent
Times zero's are calibrated to minimize the residual between the measured drift time and the drift time estimated from the tracking per wire.
The measured drift time is formulated as follows:
\begin{equation}
T_{drift} = T_0-T_{evt}-T_{tof}-T_{prop}-T_{tw}-a\cdot {\rm TDC},
\end{equation}

\begin{figure}[!htb]
\begin{minipage}[c]{0.56\linewidth}
where  $T_{tof}$, $T_{prop}$ and  $T_{tw}$ are the flight time of the particle from the reference plane, the propagation delay along the sense wire, 
and the fluctuation due to time walk effect, respectively.
The conversion factor $a = 0.98$ nsec/count.
The $T_{evt}$ is the event-by-event fluctuation from the nominal case determined by the tracking algorithm.
Time walk's are calibrated by assuming the effect is proportional to $1/\sqrt{{\rm ADC}}$ for 299 front-end boards.
The XT function is parametrized as a fifth order Chebychev polynomials and a linear function around the cell boundary.
By taking the dependence concerning the layer, left/right passage, incident angle ($\alpha$), and polar angle ($\theta$) of the particle into account,
we determine XT's for each layer, left/right passage separately with proper ($\alpha \times \theta$) categorization. 
The spatial resolution is also determined as a function of drift length for each layer, left/right passage, and ($\alpha \times \theta$) category
shown in Fig. \ref{fig:pos.reso}.
\end{minipage}
\
\begin{minipage}[c]{0.40\linewidth}
\includegraphics[width=\linewidth]{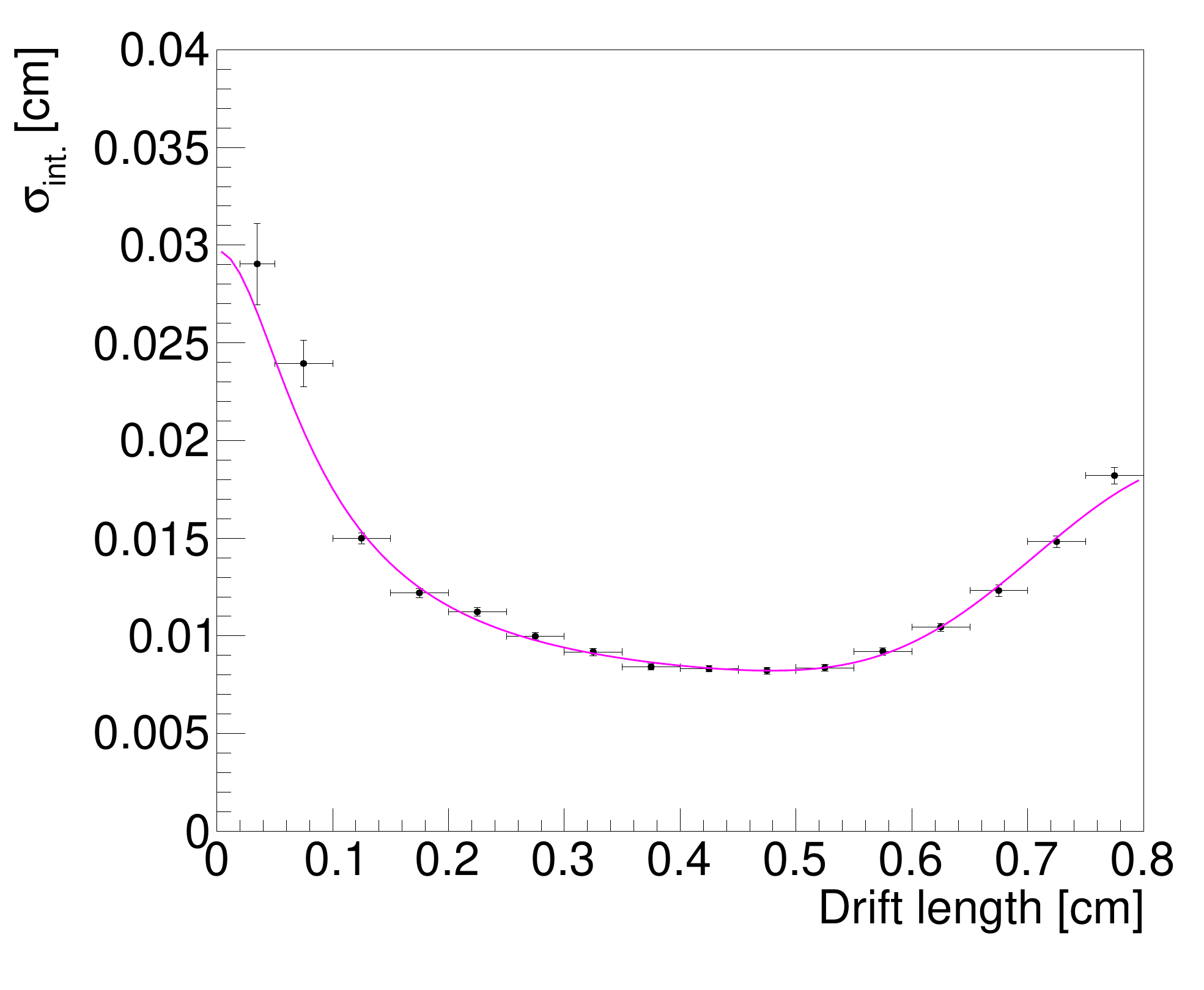}
\caption{Spatial resolution as a function of drift length.}
\label{fig:pos.reso}
\end{minipage}
\end{figure}

\begin{figure}[!htb]
\begin{minipage}[c]{0.56\linewidth}
To validate the performance of CDC after alignment  and calibration, we processed a partial sample of cosmic data with magnetic field.
The reconstructed track was separated to two parts, where one is a track reconstructed with upper half CDC, the other is one obtained with lower half of CDC.
The resolution of transverse momentum is estimated by the following equation
\begin{equation}
\frac{\sigma(P_{t})}{P_{t}} = \sqrt{2}\frac{P_t^{up} - P_t^{low}}{P_t^{up} + P_t^{low}},
\end{equation}
where $P_t^{up (low)}$ means the transverse momentum of the upper (lower) track. By fitting results shown in Figure \ref{fig:pt.reso},  $\sigma(P_{t})/P_{t}  = (0.132\pm 0.005) P_t \oplus (0.331\pm 0.016)$ was obtained.
\end{minipage}
\
\begin{minipage}[c]{0.40\linewidth}
\centering
\includegraphics[width=\linewidth, trim={2mm, 3mm, 6mm, 4mm}, clip]{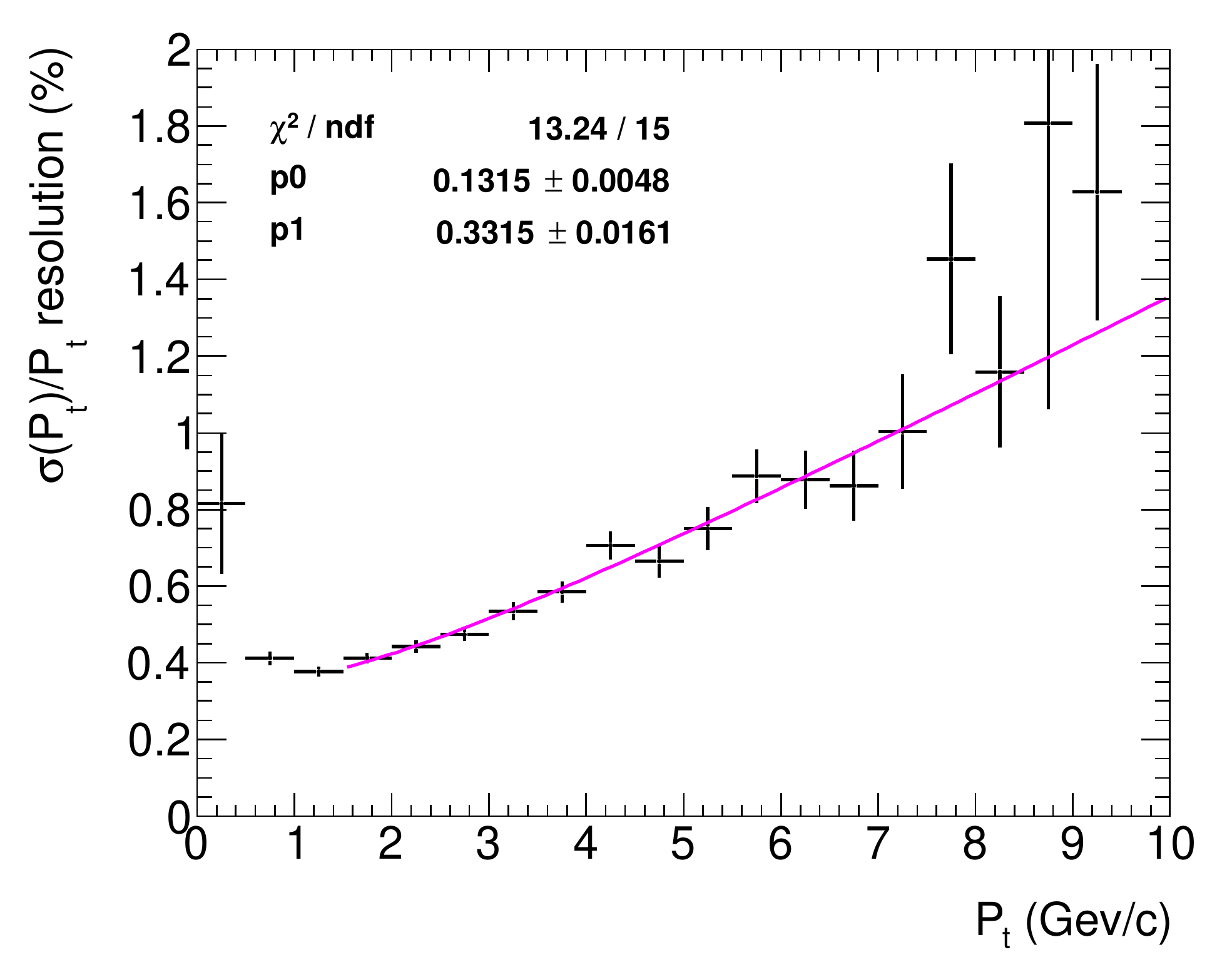}
\caption{$P_t$ resolution as a function of $P_t$}
\label{fig:pt.reso}
\end{minipage}
\end{figure}

\vspace{-2mm} 

 

%




\section{Conclusions}
Alignment and calibration constants for the Belle II tracker were determined during data taking in 2018. We extended the alignment procedure by determination of surface deformation parameters, significantly improving track-to-hit residual distributions for individual sensors and demonstrated the improvements in this paper. Quality of data was monitored and constants were validated, also as function of time. The Belle II tracker provides precise and stable measurement for time-dependent CP violation studies.





\end{document}